\documentclass[prl,aps,twocolumn,superscriptaddress]{revtex4-1}
\usepackage{graphicx,color}
\usepackage{amsthm}
\usepackage{amsfonts}
\usepackage{algorithmic}
\usepackage{enumerate}
\usepackage{latexsym}
\usepackage{amsmath}
\usepackage{amssymb}
\usepackage[colorlinks=true,citecolor=blue,linkcolor=blue]{hyperref}

\emergencystretch=\maxdimen
\begin{document}
\title{ $\textit{Zitterbewegung}$ near new Dirac points in graphene superlattice}
\author{Jianli Luan}
\affiliation{Department of Physics, Beijing Normal University, Beijing
100875, China\\}
\author{Shangyang Li}
\affiliation{Department of Physics, Beijing Normal University, Beijing
100875, China\\}
\author{Tianxing Ma}
\email{txma@bnu.edu.cn}
\affiliation{Department of Physics, Beijing Normal University, Beijing
100875, China\\}
\affiliation{Beijing Computational Science Research Center, Beijing 100193, China}
\author{Li-Gang Wang}
\email{sxwlg@yahoo.com}
\affiliation{Department of Physics, Zhejiang University, Hangzhou
310027, China\\}

\date{\today}
\begin{abstract}
New Dirac points appear when periodic potentials are applied to graphene, and there are many interesting effects near these new Dirac points. Here we investigate the $\textit{Zitterbewegung}$ effect of fermions described by a Gaussian wave packet in graphene superlattice near new Dirac points. The $\textit{Zitterbewegung}$ near different Dirac points has similar characteristics, while Fermions near new Dirac points have different group velocities in both $x$- and $y$-direction, which causes the different properties of the $\textit{Zitterbewegung}$ near new Dirac points. We also investigate the $\textit{Zitterbewegung}$ effect influenced by all Dirac points, and get the evolution with changing potential. Our intensive results suggest that graphene superlattice may provide an appropriate system to study $\textit{Zitterbewegung}$ effect near new Dirac points experimentally.

\end{abstract}

\pacs{Valid PACS appear here}
\maketitle


\section{\label{sec:level1}Introduction}
$\textit{Zitterbewegung}$ (ZB), put forward by Schr\"{o}dinger\cite{Schr1930}, is a theoretical prediction of relativistic quantum mechanics, which is a high frequency trembling motion of the electron in vacuum and resulted from the the interference between the positive and
negative energy states. Since the high characteristic frequency, it is still difficult to observe ZB of the electron in free space through experiments.

Interestingly, the concept of ZB can also be applied in some non-relativistic system, because of the similarity between the relativistic electron and two interacting bands in a solid\cite{Ferrari1990,Cannata1990}. ZB phenomenon has been predicted theoretically in variety of systems, such as semiconductors\cite{Zawadzki2005}, superconductors\cite{Cserti2006}, carbon nanotubes\cite{Zawadzki2006,Rusin2014}, topological insulators\cite{Shi2013}, some optical systems\cite{Zhang2008,Wang2009,Li2016}, ultracold atoms\cite{Zhang2013}, Weyl semimetals\cite{Huang2017}, MoS$_2$\cite{Singh2014}, graphene
\cite{Shi2013,Novoselov2005,Rusin2007,Maksimova2008,Wang2014,
Ghosh2015,Romera2009,Garcia2014,Rusin2008},
as well as the spin-orbit coupling material\cite{Shi2013,Biswas2012,*Biswas2015,*Biswas2017}. Shi et. al. provided a general method to derive the analytical expression for the ZB effect then use it to investigate ZB in many different systems\cite{Shi2013}.
There is also an experimental observation of ZB using Bose-Einstein condensate\cite{LeBlanc2013}.
Although it is difficult to observe ZB of electron in graphene because of its high frequency and low amplitude, the ZB phenomenon may be observed under current experiment conditions when a 1D periodic potential applies to it\cite{Wang2014}. For this reason,
it is 
significant to investigate ZB in 1D periodic graphene superlattice.

In this paper, we investigate the ZB phenomenon near new Dirac points of fermions in graphene superlattice. By applying a periodic potential to monolayer graphene, its band structure will change, and in some specific conditions, new Dirac points will appear\cite{Park20081,Park20082,Park2009,Brey2009,Barbier2010,Wang2010}.
Few works has been focused on ZB near these new Dirac points.
When a periodic potential applies to graphene,
the group velocity of the fermions turns to be highly anisotropic,
and it has been illustrated that the $y$-direction group velocity will decrease but $x$-direction contain unchanged near original Dirac point.
Nevertheless, near new Dirac points, both $x$- and $y$-direction group velocities will decrease\cite{Park2009,Brey2009,Barbier2010},
and then influence ZB effect remarkably.

In the following, we concentrate on the different features of ZB when the fermions are near different Dirac points and described by a Gaussian wave packet, which are influenced by group velocities, wave packets and distance between the fermion and Dirac points. It is demonstrated that ZB has similar features and change rules near different Dirac points, but original Dirac point is special because of the symmetry and unchanged $x$-direction group velocity.
The decrease of group velocities in two directions will affect ZB remarkably. These influences depend on the oscillation direction in real space and initial momentum of wave packets.
The evolution of ZB with changing initial momentum of fermions will be more complicatedly when there are new Dirac points. By investigating the ZB controlled by different Dirac points, we find that these oscillations may have the most prominent one when the Dirac points are far away, or be similar when the Dirac points are close, which can be regarded as controlled by all Dirac points. In addition, we study the evolution of ZB when the amplitude of periodic potential and the number of Dirac points are changing. It is indicated that the condition for obtaining ZB controlled by all Dirac points is that we should choose appropriate potential and wave packet to make all Dirac points inside it. Therefore, the ZB oscillation is controlled by the nearest Dirac point mainly in most instances and graphene superlattices may provide an appropriate system to research ZB near new Dirac point experimentally.

\section{\label{sec:level2}Model and method}
We consider the low-energy electronic states of graphene, the Hamiltonian can be written as\cite{Wallace1947}
\begin{equation}
H=
\hbar v_f(k_x\sigma_x+k_y\sigma_y),\label{eq1}
\end{equation}
where $\mathit{v_f}$ is Fermi velocity $10^6m/s$, $\sigma_x$ and $\sigma_y$ are pauli matrices and $\vec{k}$ is the wave vector from $K$ point. For monolayer graphene, the position operator can be calculated by substituting Eq. (\ref{eq1}) into $x(t)=e^{i\frac{H}{\hbar}t}x(0)e^{-i\frac{H}{\hbar}t}$ and its (1,1) component is
\begin{equation}
{x}_{11}(t)=x_{11}(0)+\frac{k_y}{2k^2}[1-\cos{(2v_fkt)}],\label{eq2}
\end{equation}

\begin{equation}
{y}_{11}(t)=y_{11}(0)-\frac{k_x}{2k^2}[1-\cos{(2v_fkt)}].\label{eq22}
\end{equation}
These equations can also be obtained by using general formula established by\cite{Shi2013}. It has also illustrated that ZB is controlled by effective Fermi velocity and the distance between fermions and Dirac point in previous research\cite{Zhang2008,Huang2017,Rusin2007,Wang2014}.

\begin{figure}[tb]
  \centering
  \includegraphics[width=8.5cm]{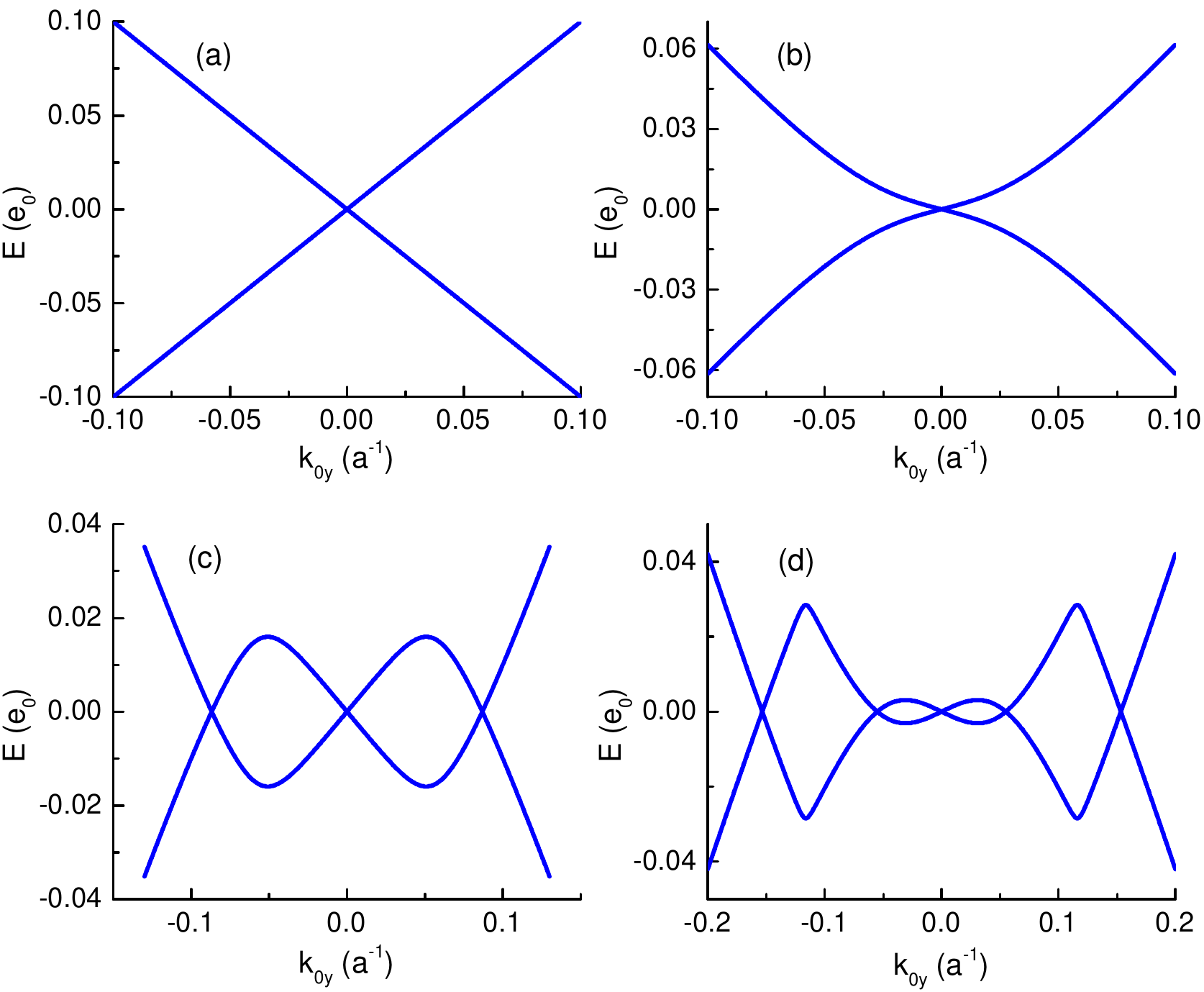}
  \caption{(Color online) Band structure of graphene (a) and grephene superlattice with different $V_0$, which $V_0=n\pi\frac{\hbar v_f}{L}$ and (b) (c) (d) are n=2, 4, 6 respectively. Energy is in units of $e_0=\frac{\hbar v_f}{L}$ and $k_y$ is in units of $a^{-1}$, which $a$ is lattice constant of graphene and $L=100a$. It demonstrates that new Dirac points appear in specific potential. All new Dirac points are symmetric about the original one. }\label{fig1}
\end{figure}

We further consider a one dimensional periodic potential along the $\mathit{x}$ direction $\mathit{V(x)=V_0\cos G_0x}$, which is applied to graphene, with the periodicity is $\mathit{L}$ and $G_0=2\pi/\mathit{L}$, so the total Hamiltonian can read
\begin{equation}
H=
\hbar v_f(k_x\sigma_x+k_y\sigma_y)+V(x)I,\label{eq3}
\end{equation}
where $\mathit{I}$ is a 2$\times$2 unit matrix. We take the form of $\mathit{V_0}$ as $\mathit{V_0=n\pi\frac{\hbar v_f}{L}}$. With different $\mathit{n}$, the number of Dirac points in band structure is different, since new Dirac points appear in some $\mathit{V_0}$. When $n$ is a root of equation $J_0(\frac{2V_0}{\hbar v_fG_0})=0$, which $J_0$ is 0th Bessel function of the first kind, a pair of new Dirac points will appear in band structure\cite{Park2009}. The Fig. \ref{fig1} shows band structure in different $\mathit{n}$, We can see that the number of Dirac points is different in Fig. \ref{fig1}, and new Dirac points are symmetric about the original Dirac point. We mark the Dirac points from the original point of the coordinate system with 0, 1, 2...for convenience.

One of the most important influences of periodic potential is that it can decrease the group velocities of electrons in real space, which can influence ZB oscillation.
When a periodic potential is applied to monolayer graphene, the $y$-direction group velocity of fermion near original Dirac point will decrease remarkably, but its $x$-direction group velocity will not change. Differently, near new Dirac points, fermion's $x$- and $y$-direction group velocity will both decrease\cite{Barbier2010,Park2009}.
We use $f_x$ and $f_y$ to indicate the ratio of $x$- and $y$-direction group velocity and $v_f$ respectively ($v_x=f_xv_f$, $v_y=f_yv_f$), and they can be obtained by calculating the slope of band structure in momentum space\cite{Barbier2010}.
In particularly, for cosine potential, $f_y=J_0(\frac{2V_0}{\hbar v_fG_0})$ when fermion is near original Dirac point in first Brillouin zone\cite{Brey2009}.

We use the same method and substitute Eq. (\ref{eq3}) into $x(t)=e^{i\frac{H}{\hbar}t}x(0)e^{-i\frac{H}{\hbar}t}$  to get (1,1) component of position operator which can be given as
\begin{equation}
{x}_{11}(t)=x_{11}(0)+\frac{k_y}{2k^2}[1-\cos{(2v_fkt)}],\label{eq4}
\end{equation}

\begin{equation}
{y}_{11}(t)=y_{11}(0)-\frac{k_x}{2k^2}[1-\cos{(2v_fkt)}],\label{eq44}
\end{equation}
which $k=\sqrt{k_x^2+k_y^2}$. They have the same form as Eq. (\ref{eq2}) and (\ref{eq22}), but in this equation, effective Fermi velocity $v_f$ and location of Dirac points, which decides wave vector from $K$ point $k_y$ and the distance between Dirac point and fermion, are different when periodic potential applies to graphene. So we rewrite Eq. (\ref{eq4}) and (\ref{eq44}) to make periodic potential's influences clearer.
\begin{equation}
{x}_{11}(t)=x_{11}(0)+\frac{f_xf_y\kappa_y}{2\kappa^2}[1-\cos{(2v_f\kappa t)}],
\end{equation}

\begin{equation}
{y}_{11}(t)=y_{11}(0)-\frac{f_xf_y\kappa_x}{2\kappa^2}[1-\cos{(2v_f\kappa t)}],
\end{equation}
where $\kappa=\sqrt{f_x^2\kappa_x^2+f_y^2\kappa_y^2}$,
$\kappa_x=k_x-k_{Dx}$, $\kappa_y=k_y-k_{Dy}$, which ($k_{Dx}$,$k_{Dy}$)
is the location of Dirac point in momentum space and $\kappa_x$ and $\kappa_y$ indicates the wave vector from Dirac point, which can also represent the distance between fermion and Dirac point and are different from $k_x$ and $k_y$ respectively because new Dirac points are not at original point anymore. $f_x$, $f_y$ indicate change of $x$- and $y$-direction effective Fermi velocity respectively and $v_f$ is $10^6m/s$, so we can get $v_x=f_xv_f$, $v_y=f_yv_f$ as group velocity of fermions along $x$- and $y$-direction respectively. We can find that the motion of the fermion consists of an oscillation with frequency $2v_f\sqrt{f_x^2\kappa_x^2+f_y^2\kappa_y^2}$.

We consider the condition that the initial state of Dirac fermions can be described by a Gaussian wave packet\cite{Schliemann2006,Schliemann2005}
\begin{equation}
\psi(\vec{r},0)=\frac{d}{2\pi ^{3/2}}\int d^2\vec{k}e^{-\frac{1}{2}d^2(k_x-k_{0x})^2-\frac{1}{2}d^2(k_y-k_{0y})^2}e^{i\vec{k}\cdot \vec{r}}\left(\begin{array}{c}1\\0\end{array} \right),
\end{equation}
which $\mathit{d}$ is the width of the packet, and ($k_{0x}$,$\mathit{k_{0y}}$) is the center of the packet in momentum space. For simplicity, if we do not indicate $k_{0x}$, it will be $0$. The unit vector (1,0) is a convenient choice\cite{Schliemann2005}, and then the average of the (1,1) component of $\mathit{x(t)}$
and $y(t)$
can be written as
\begin{equation}
\begin{aligned}
\bar{x}_{11}(t)&=\langle{\psi}|{x}_{11}(t)|{\psi}\rangle\\
&=\frac{d^2}{\pi}\iint\frac{f_xf_y\kappa_y}{2\kappa^2}[1-\cos{(2v_f\kappa t)}]\\
&\times e^{-d^2(k_x-k_{0x})^2-d^2(k_y-k_{0y})^2}dk_xdk_y,\label{eqx}
\end{aligned}
\end{equation}

\begin{equation}
\begin{aligned}
\bar{y}_{11}(t)&=\langle{\psi}|{y}_{11}(t)|{\psi}\rangle\\
&=\frac{d^2}{\pi}\iint-\frac{f_xf_y\kappa_x}{2\kappa^2}[1-\cos{(2v_f\kappa t)}]\\
&\times e^{-d^2(k_x-k_{0x})^2-d^2(k_y-k_{0y})^2}dk_xdk_y.\label{eqy}
\end{aligned}
\end{equation}

We should take an appropriate width of wave packet to research each Dirac point's influences on ZB or ZB controlled by all Dirac points commonly. Firstly, we take a big $d$ to make wave packet's width not too large in momentum space to avoid ZB being influenced by all Dirac points commonly. Then we take an appropriate $d$ to ensure that all Dirac points can be inside a wave packet in momentum space, so we can find laws of ZB controlled by all Dirac points.

The group velocities of fermions need to be considered. Near original Dirac point, we take $v_x=v_f$ and $v_y=f_yv_f=|J_0(\frac{2V_0}{\hbar v_fG_0})|v_f$\cite{Brey2009}, and these velocities are not changed along with the motion of fermions since that band structure is linear about $k_y$ near original Dirac point. Near 1st new Dirac points, we do not have a equation to calculate $f_x$ and $f_y$, and they will change along with the motion of fermion, since band structure is not linear anymore. So we get them by calculating the $k_x$- and $k_y$-direction slope at the location of wave packet's center in band structure as $f_x$ and $f_y$ respectively. Since other periodic potential can also change fermions' group velocities, it can also influence ZB oscillation by changing group velocities, and our model can also be used to research other periodic potential.
We also notice that both two directions' group velocities will change when potentials are applied to graphene in two directions\cite{Park20082}. Our method can also be used for this situation and may obtain similar properties.

\section{\label{sec:level3}results and discussion}

\begin{figure}[tb]
  \centering
  \includegraphics[width=8.5cm]{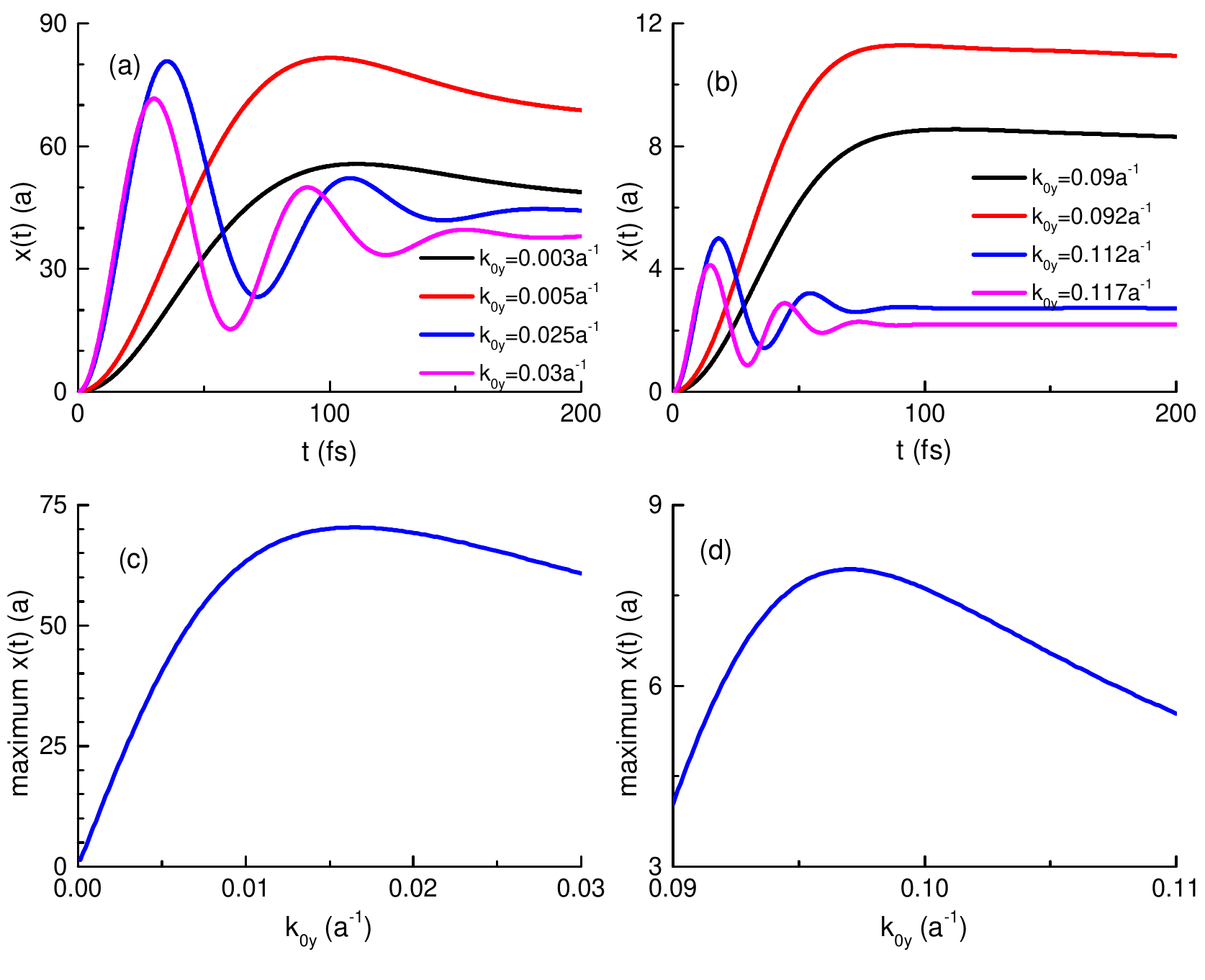}
  \caption{(Color online) The $x$-direction ZB oscillation near original Dirac point (a) and 1st new one (b) with different $k_{0y}$ but the same distance from wave packet's center to Dirac point when $V_0=4\pi\frac{\hbar v_f}{L}$. (c) (d) are maximum amplitude as a function of $k_{0y}$ when $V_0=4\pi\frac{\hbar v_f}{L}$. The band structure is shown in Fig. \ref{fig1}(c) and here $L=100a$. In (a) (b), $f_x$ and $f_y$ are calculated in the way mentioned in model and $d=150a$.
  In (c) (d), $f_y=0.397$, $f_y=0.75$ respectively and $d=100a$.}\label{fig2}
\end{figure}

We first study ZB controlled by each Dirac point independently. We take $V_0=4\pi\frac{\hbar v_f}{L}$, which $L=100a$ and $a$ is lattice constant of graphene. The band structure is shown in Fig. \ref{fig1}(c) and it is demonstrated that new Dirac points locate at $k_y=-0.087a^{-1}$ and $k_y=0.087a^{-1}$ in momentum space. We take the width of wave packet $d=150a$. Different potentials mainly influence ZB by changing group velocities of fermions, which is reflected in $f_x$ and $f_y$ and we can get them in the way mentioned in model.
In this potential, $x$-direction group velocity is not changed and $f_y=|{J_0(4)}|=0.397$ near original Dirac point. Near 1st new Dirac point, we measure the $k_x$- and $k_y$-direction slope of band structure in momentum space at wave packet's center as $f_x$ and $f_y$ respectively. Since $f_x$ changes slowly, we regard it as a constant near the same Dirac point.

We make the center of wave packets have the same distance to Dirac points in momentum space, and it is reasonable that ZB has similar properties near different points, which is demonstrated in Fig. {\ref{fig2}}(a) and (b). In these two figures, the oscillations have similar shapes, but different period, amplitude and attenuation, since fermions near new Dirac points has different group velocities, affecting the period and amplitude of oscillations. We can also find that with increasing $k_{0y}$, the amplitude of oscillation increase firstly then decrease, and the period decrease all the time near both original and new Dirac points. It has also demonstrated that decreasing $k_{0y}$ leads to slower attenuation. We show the relation of $k_{0y}$ and maximum amplitude in Fig. \ref{fig2}(c) and (d) (In these figures, we assume $f_x$ and $f_y$ are constant to find the relation of amplitude and $k_{0y}$). We can find this property in these two figures.
Since ZB represents oscillation caused by interference between positive and negative energy state, the smaller energy gap will cause this interference more easily then make ZB have slower attenuation, larger amplitude and period. It is also shown in Eq. (\ref{eqx}) and (\ref{eqy}) that the frequency is nearly proportional to $k_{0y}$. However, in Eq. (\ref{eqx}) and (\ref{eqy}), the amplitude is nearly proportional to $\kappa_x$ or $\kappa_y$, which represents the distance between wave packet and Dirac points. For this reason, ZB cannot appear when wave packet is at Dirac point in momentum space. Therefore, with increasing $k_{0y}$, the energy gap increase then cause smaller period and quicker attenuation. For amplitude, ZB will arise firstly with the increase of the distance between wave packet and Dirac point, then will decrease because of the increasing energy gap.

It is also shown that the amplitude and period change more remarkably with increasing $k_{0y}$ in Fig. \ref{fig2}(b), since near new Dirac point, group velocities change with different $k_{0y}$, but contain a constant near original Dirac point.
The change of group velocities will change the shape of band structure then affect the energy gap and ZB, which will be discussed in detail later.
Near original Dirac point, its $v_x=v_f$, $v_y=f_yv_f$ and $f_y$ is constant, so the oscillation will evolve according to the rule reflected by Fig. \ref{fig2}(c) with different $k_{0y}$ in momentum space. Near 1st Dirac points, when the wave packet center is moving away from it, $k_{0y}$ will increase and group velocities will also change, so the oscillation's evolution is more complicated. It is influenced by the $k_{0y}$ and group velocities simultaneously, not changes as Fig. \ref{fig2}(d) simply.

\begin{figure}[tb]
  \centering
  \includegraphics[width=8.5cm]{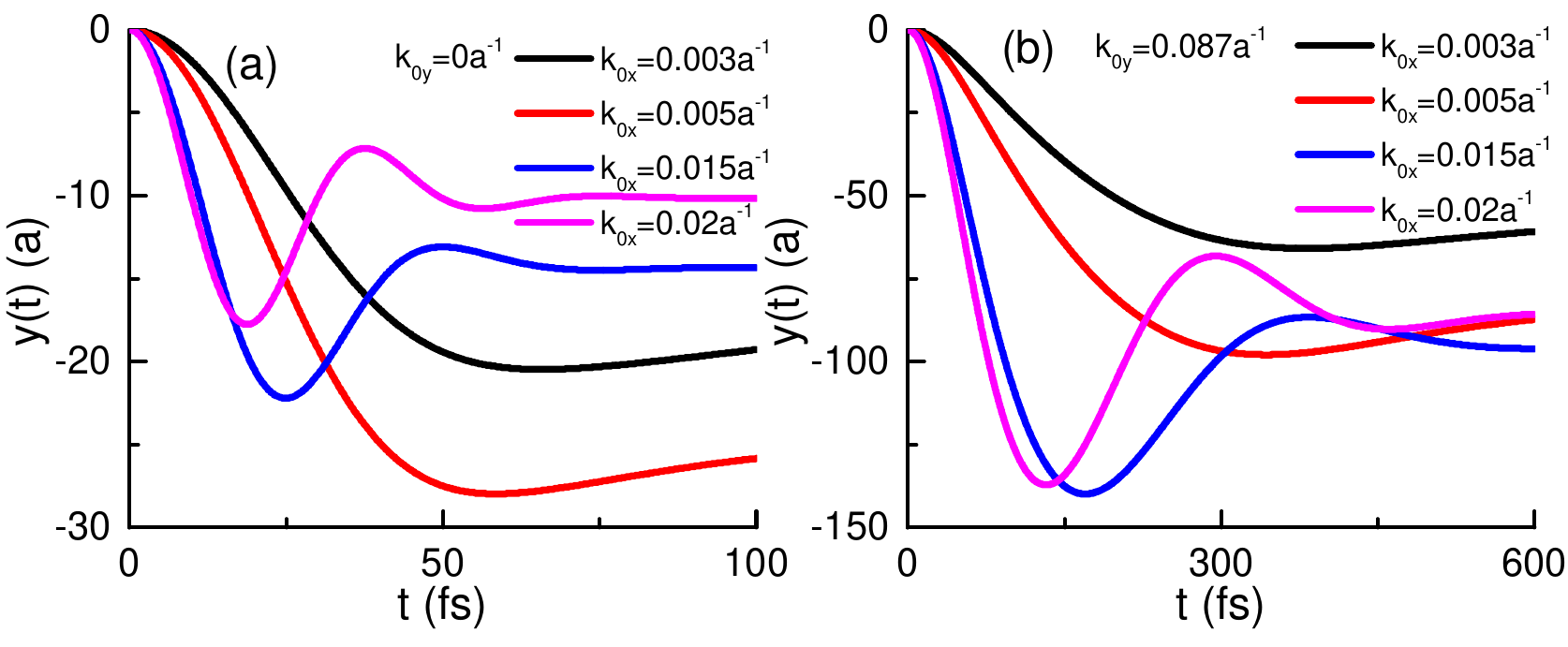}
  \caption{(Color online) The $y$-direction ZB oscillation near original Dirac point (a) and 1st new one (b) with different $k_{0x}$ but the same distance from wave packet's center to Dirac point when $V_0=4\pi\frac{\hbar v_f}{L}$. Other parameters are the same as Fig. \ref{fig2}(a) and (b).}\label{fig3}
\end{figure}

In Fig. \ref{fig3}, we also provide $y$-direction oscillation in real space, but with different $k_{0x}$ and the same $k_{0y}$ in momentum space. With the increase of $k_{0x}$, it is reasonable that the amplitude, period and attenuation of ZB have similar properties which are shown in Fig. \ref{fig2} for the same reason discussed above. However, near 1st new Dirac point, $y$-direction ZB oscillation has larger amplitude and period, which is different from $x$-direction and beneficial to be observed. Therefore, to observe ZB near new Dirac points more easily in experiments, one can choose $y$-direction oscillation with $k_x$-direction initial momentum.

\begin{figure}[tb]
  \centering
  \includegraphics[width=8.5cm]{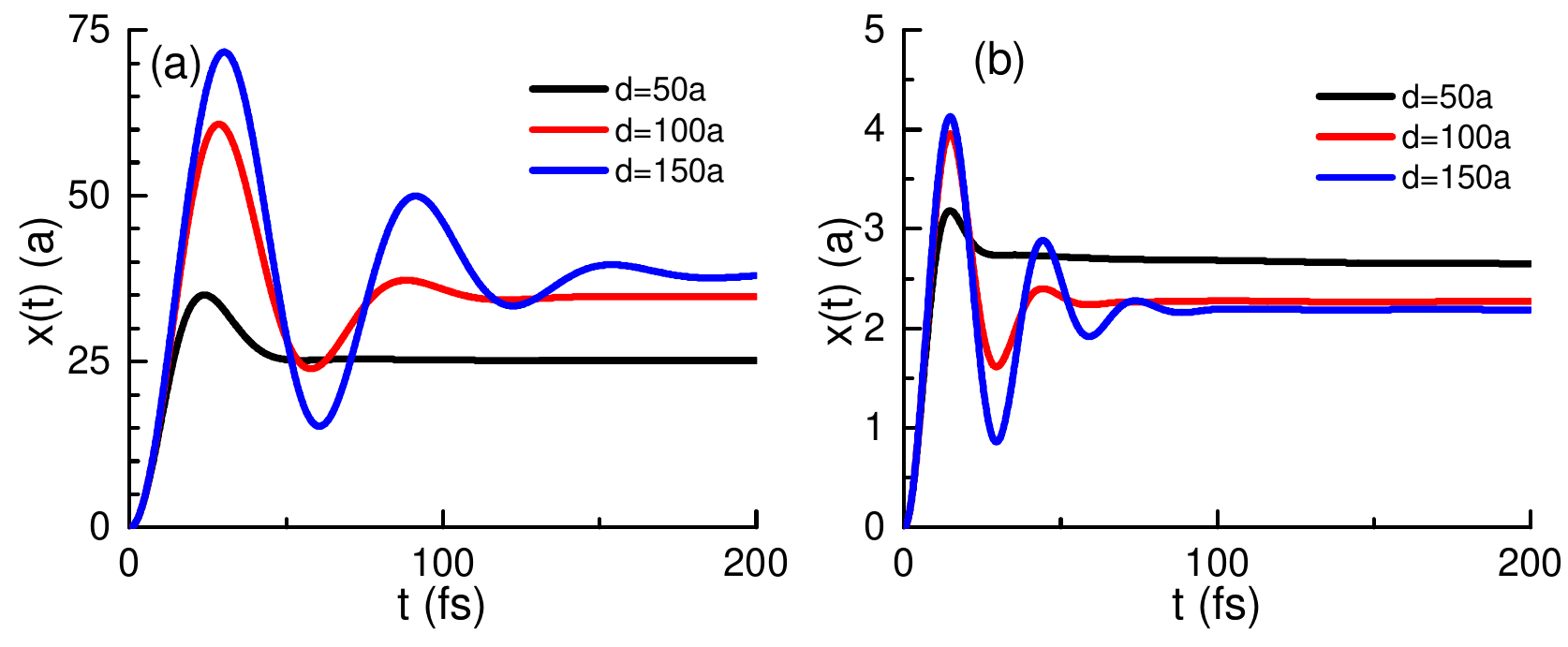}
  \caption{(Color online) The ZB oscillation with different width of wave packets. (a) shows the evolution of wave packet near original Dirac point with $k_{0y}=0.03a^{-1}$, and (b) is near 1st right new Dirac point with $k_{0y}=0.117a^{-1}$. Other parameters are the same as Fig. \ref{fig2}(a) and (b).}\label{fig4}
\end{figure}

The ZB oscillation is influenced by the width of wave packet. In Fig. \ref{fig4}, we take $k_{0y}=0.03a^{-1}$ and $k_{0y}=0.117a^{-1}$ in momentum space, which is near original and 1st right new Dirac point respectively, and different $d$. It can be concluded from Fig. \ref{fig4} that ZB has increasing amplitude with increasing $d$ near both original and new Dirac points, which can be observed more easily in experiments.
Although period also increases a little with $d$, it almost maintains constant. This is consistent with the previous research about original Dirac point\cite{Zhang2008,Rusin2007,Wang2014}, which can also reflect the similarities of the ZB oscillation near different Dirac points.
From Eq. (\ref{eqx}), we can find that amplitude is proportional to $d^2$ and period dependents weakly on $d$. It also contains an exponential decay component, which is dependent strongly on $d$, then causes the transient characteristic and slower decay with larger $d$.

\begin{figure}[tb]
  \centering
  \includegraphics[width=8.5cm]{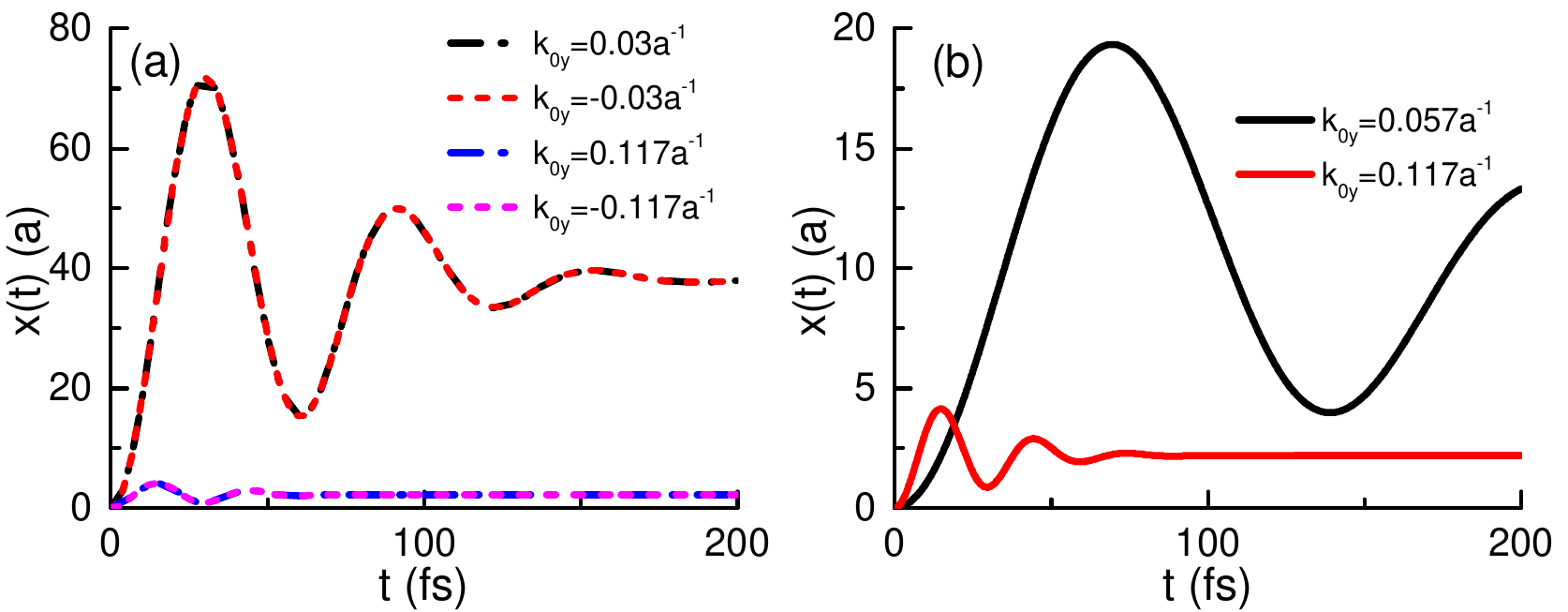}
  \caption{(Color online) The ZB oscillation when wave packets are symmetric about original Dirac point (a) and 1st right new one (b). Except for $k_{0y}$, all the other parameters are the same as Fig. \ref{fig2}(a) and (b).}\label{fig5}
\end{figure}

The symmetry of the band structure's influences on ZB are studied. In Fig. \ref{fig5} we take wave packets that their centers are symmetric about original or new Dirac points in momentum space. From Fig. \ref{fig5}(a), we can find that the characteristics of oscillations are the same when the wave packet center is in two positions that are symmetric about the original Dirac point. On the contrary, in two symmetric locations about the 1st right new Dirac point, because of the different $k_{0y}$ and slope, we can find amplitude and period are different in Fig. \ref{fig5}(b). We also notice that when $k_{0y}=0.057a^{-1}$ in Fig. \ref{fig5}(b), the oscillation has large period, since band structure here has a small slope, which make fermions have small group velocities. In this condition, the ZB oscillation is affected by original Dirac point, so we can conclude that ZB will controlled by both new and original Dirac points in some specific condition.

\begin{figure}[tb]
  \centering
  \includegraphics[width=8.5cm]{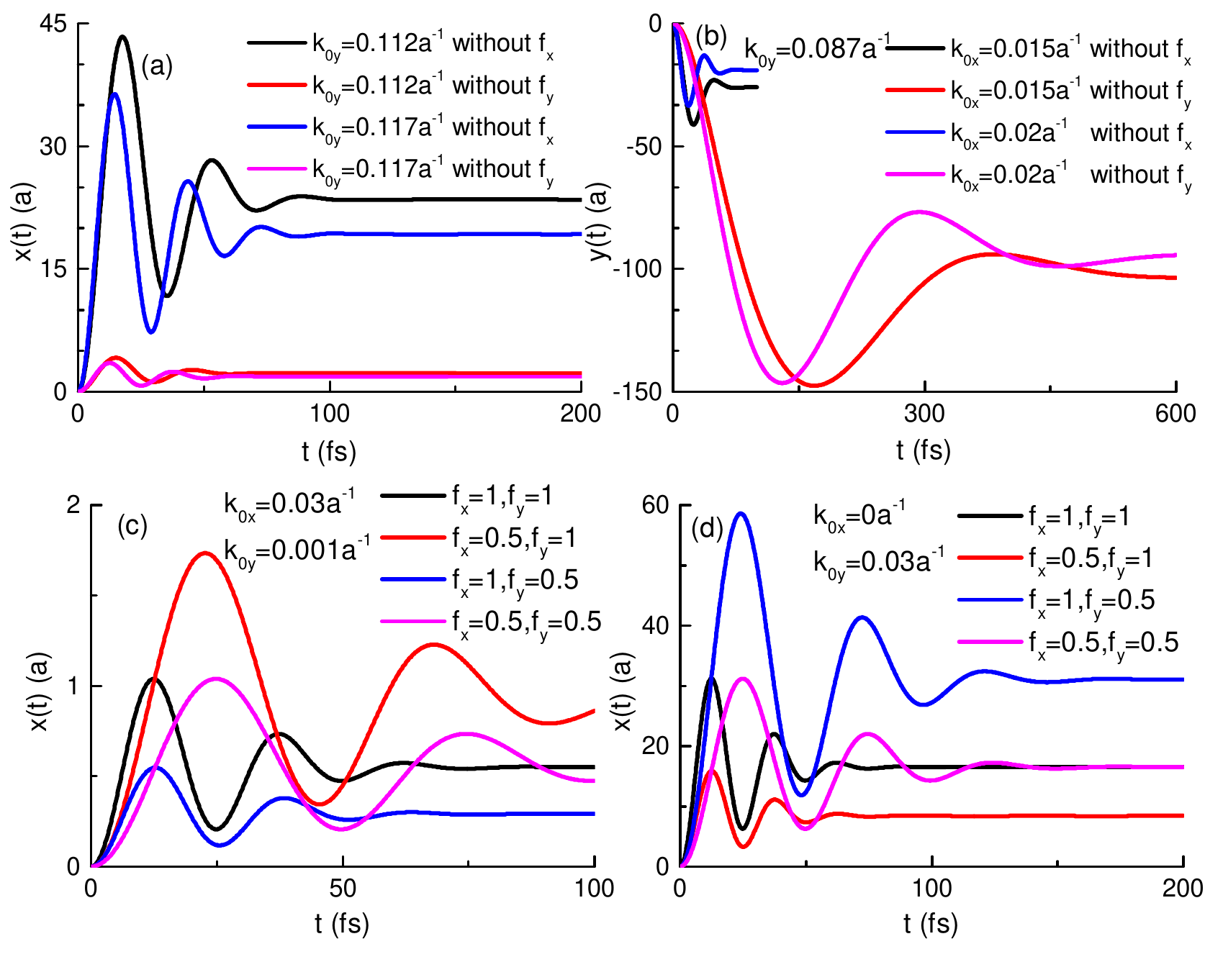}
  \caption{(Color online) The $x$-(a) and $y$-direction(b) ZB oscillation without $f_x$ or without $f_y$. (c) and (d) are ZB oscillations with some specific $f_x$ and $f_y$ and wave packets move along $k_x$- or $k_y$-direction in momentum space respectively. Other parameters are the same as Fig. \ref{fig2}(b).}\label{fig6}
\end{figure}

Previous works illustrated that $f_y$ increases the amplitude and period of ZB when fermions are near original Dirac point and wave packets move along $k_y$- direction in momentum space.\cite{Wang2014}.
In our research, fermions' $x$-direction velocities will also change, and initial momentum may be in $k_x$ direction.
In Fig. {\ref{fig6}}(a)(b), we show ZB without $f_x$ or without $f_y$ when fermions are near new Dirac point.
To understand effects of $f_x$ and $f_y$ more explicitly, we also set some specific $f_x$ and $f_y$ and obtain ZB when wave packets move along $k_x$- and $k_y$-direction respectively.
Fig. \ref{fig6}(c) shows both $f_x$ and $f_y$ can increase period but the influence of $f_x$ is more remarkable when initial momenta are along $k_x$-direction. Opposite conclusion can be obtained from Fig. \ref{fig6}(d). This is because that energy gap depends on $x$-direction group velocity mainly if wave packets move along $k_x$-direction in momentum space. Small group velocity means more flat band structure and small energy gap, which can cause large period. Meanwhile, the frequency $2v_fk$ is proportional to group velocity in Eq. (\ref{eqx}) and (\ref{eqy}), so the decrease of two directions' group velocities can both increase period.
We can also find that the oscillations have larger amplitude when $f_x<1$ in Fig. \ref{fig6}(c) than $f_x$=1, but smaller ones when $f_y<1$. Fig. \ref{fig6}(d) exhibits opposite property. This is due to that the decrease of band structure's slope in one direction leads to small gap along the same direction in momentum space, then ZB oscillation will have large amplitude if wave packet moves along this direction in momentum space. However, if $f_x$ and $f_y$ exist in Eq. (\ref{eqx}) and (\ref{eqy}) at the same time, the amplitude will be proportional to one of them, then the amplitude will decrease.
This property is a main influence of the anisotropy in graphene superlattice.

\begin{figure}[tb]
  \centering
  \includegraphics[width=8.5cm]{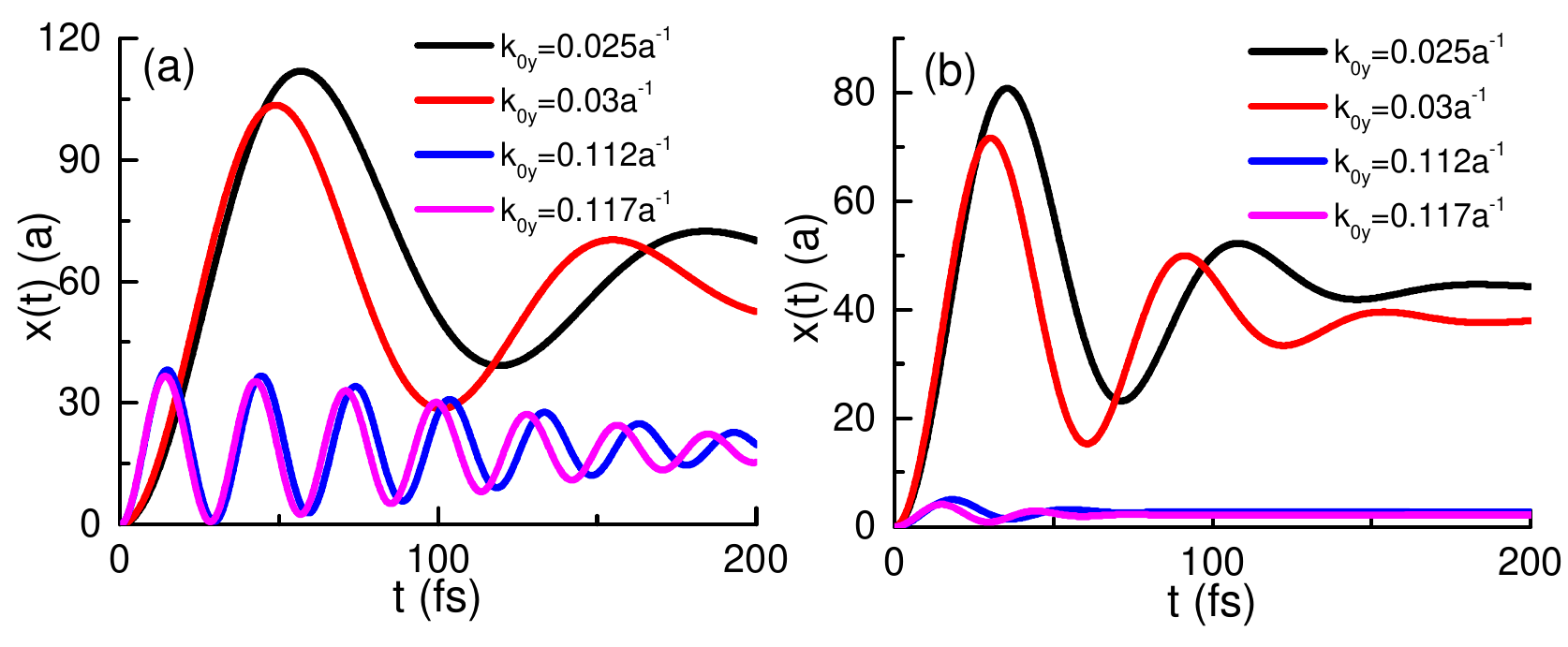}
  \caption{(Color online) The ZB oscillation when $V_0=2\pi\frac{\hbar v_f}{L}$ (a) and $V_0=4\pi\frac{\hbar v_f}{L}$(b) with the same $k_{0y}$. $f_x=1$, $f_y=0.224$ in (a) and they are gotten by calculating slope of band structure in (b). Other parameters are the same as Fig. \ref{fig2}(a) and (b).}\label{fig7}
\end{figure}

We take two $V_0$ to find the differences about ZB when there are new Dirac points or not. We take $V_0=2\pi\frac{\hbar v_f}{L}$ (without new Dirac points) and $V_0=4\pi\frac{\hbar v_f}{L}$ (with new Dirac points). In Fig. \ref{fig7}(a), we can see that the oscillations evolve as the rule shown in Fig. \ref{fig2}(c), since group velocities are unchanged. So we can get a high frequency ZB oscillation by making center of wave packet far away from Dirac point. In Fig. \ref{fig7}(b), we can find two similar oscillation at different $k_{0y}$ since there are new Dirac points. The oscillation evolution rules with changing $k_{0y}$ is not so simple anymore. If we want to observe ZB controlled by a single Dirac point, We need to choose a large $d$ in real space to avoid other Dirac points' influences in momentum space and an appropriate $k_{0y}$ to make amplitude and frequency fit experiments conditions.

\begin{figure}[tb]
  \centering
  \includegraphics[width=8.5cm]{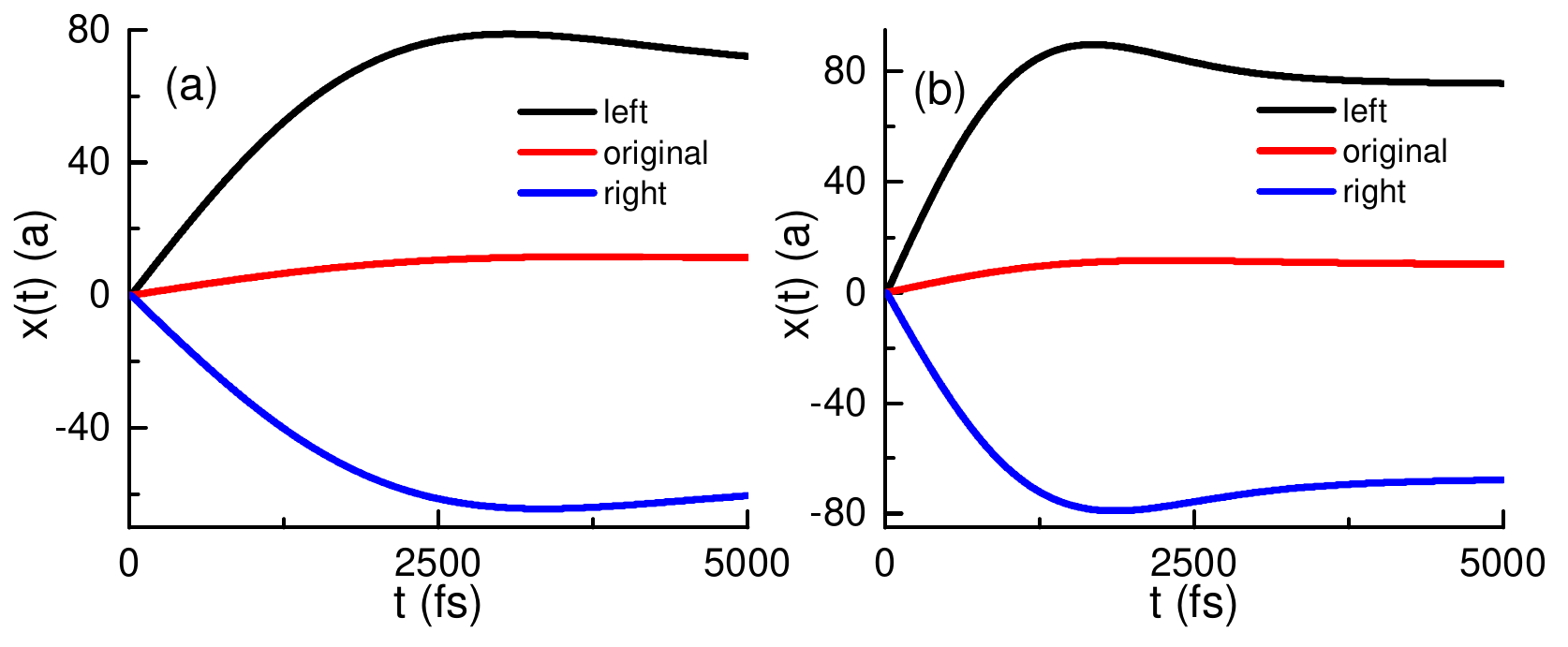}
  \caption{(Color online) The ZB oscillation when $V_0=2.42\pi\frac{\hbar v_f}{L}$ (a) and $V_0=2.43\pi\frac{\hbar v_f}{L}$ (b). Here original, right and left indicate ZB controlled by original Dirac point, 1st right and left new Dirac points respectively. $d=100a$, $k_{0y}=0.001a^{-1}$, $f_x=1$ and $f_y=0.0079 (a), 0.013 (b)$.}\label{fig8}
\end{figure}

\begin{figure}[tb]
  \centering
  \includegraphics[width=8.5cm]{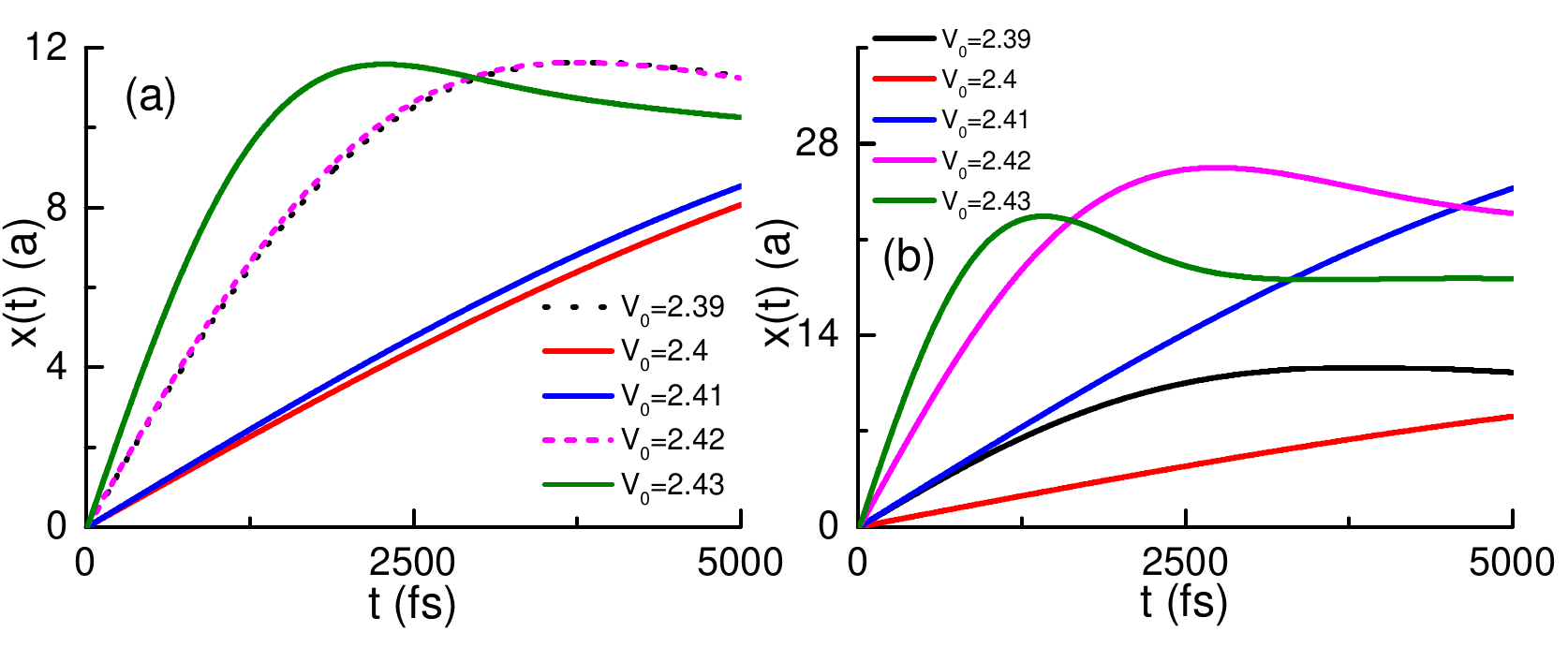}
  \caption{(Color online) The ZB oscillation controlled by original Dirac point (a) and all Dirac points (b) with different $V_0$ in units of $e_0=\pi\frac{\hbar v_f}{L}$. $d=100a$, $f_x=1$ and $f_y$ are obtained by calculating $J_0(\frac{2V_0}{\hbar v_fG_0})$. Other parameters are the same as Fig. \ref{fig8}.}\label{fig9}
\end{figure}

Secondly, we consider ZB oscillation that controlled by all Dirac points in common by researching the evolution of wave packets and ZB oscillation when $V_0$ is changing. We take an appropriate wave packet to ensure that all Dirac points are inside it. There are three Dirac points when $V_0$ are $2.405-5.52\pi\frac{\hbar v_f}{L}$\cite{Park2009}. We take $V_0=2.39, 2.40, 2.41, 2.42, 2.43\pi\frac{\hbar v_f}{L}$ to research, which can get the evolution of ZB when the number of Dirac points changes. To make all Dirac points inside wave packet, we take $d=100a$ and $k_{0y}=0.001a^{-1}$, and we can get $f_x=1$ and $f_y=J_0(\frac{2V_0}{\hbar v_fG_0})$. Since $k_{0y}$ and the slope are small, and slope change slowly, we think $f_x$ and $f_y$ are constant in this condition. We first take $V_0=2.42,2.43\pi\frac{\hbar v_f}{L}$ and research ZB oscillation controlled by three Dirac points respectively. Since all Dirac points are close to each other in momentum space, if we take a wave packet near original one, it will also be close to others, then create three oscillations, and we get oscillation curves in Fig. \ref{fig8}. We can see that all oscillations have similar period and amplitude, so we can superpose them to an oscillation, which will be discussed later.
The amplitude and frequency is small, since $k_{0y}$ and slope of band structure are too small to produce efficient interference in this condition.
Since two new Dirac points are at different side of wave packet, the oscillations have opposite direction in Fig. \ref{fig8}. Then we take five $V_0$ mentioned in the preceding part of this paragraph, and get oscillations controlled by original point (Fig. \ref{fig9}(a)) and all Dirac points (Fig. \ref{fig9}(b)). Since in Fig. \ref{fig9}(a) $f_x=1$, $f_y=J_0(\frac{2V_0}{\hbar v_fG_0})$, and $J_0$ is symmetric about 2.405, the oscillations are symmetric about $n=2.405$. On the contrary, if we superpose all oscillation, they are not symmetric anymore in Fig. \ref{fig9}(b). Moreover, when $n=2.4$ and $2.41$, new Dirac points begin to appear and the band structures are too flat, which means energy gap nearly vanish, so it is hard to obtain obvious oscillations.

\begin{figure}[tb]
  \centering
  \includegraphics[width=8.5cm]{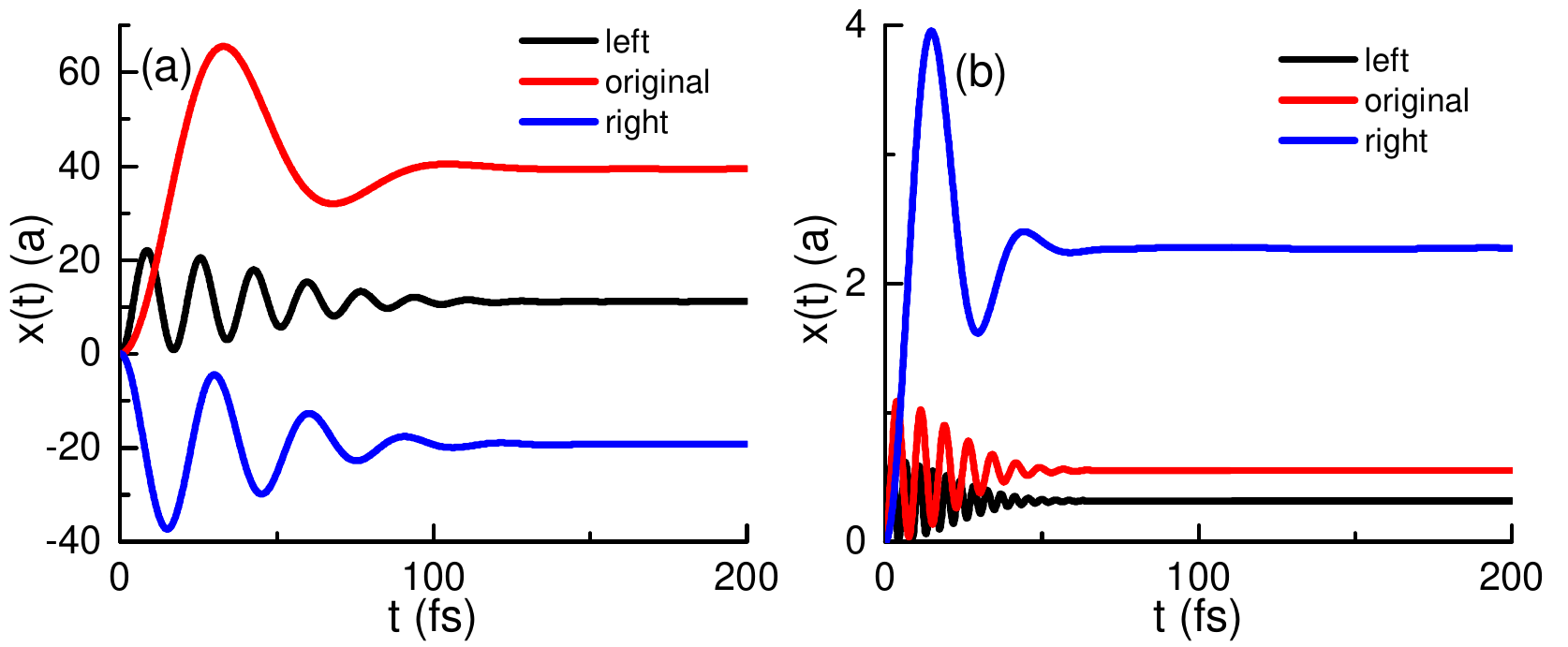}
  \caption{(Color online) The ZB oscillation when $V_0=4\pi\frac{\hbar v_f}{L}$, $k_{0y}=0.025a^{-1}$ (a) and $k_{0y}=0.117a^{-1}$ (b). Here original, right and left indicate ZB controlled by original Dirac point, 1st right and left new Dirac points respectively. $d=100a$, $f_x$ and $f_y$ can be gotten by the same way in Fig. \ref{fig2}(a) and (b).}\label{fig10}
\end{figure}

\begin{figure}[tb]
  \centering
  \includegraphics[width=4.25cm]{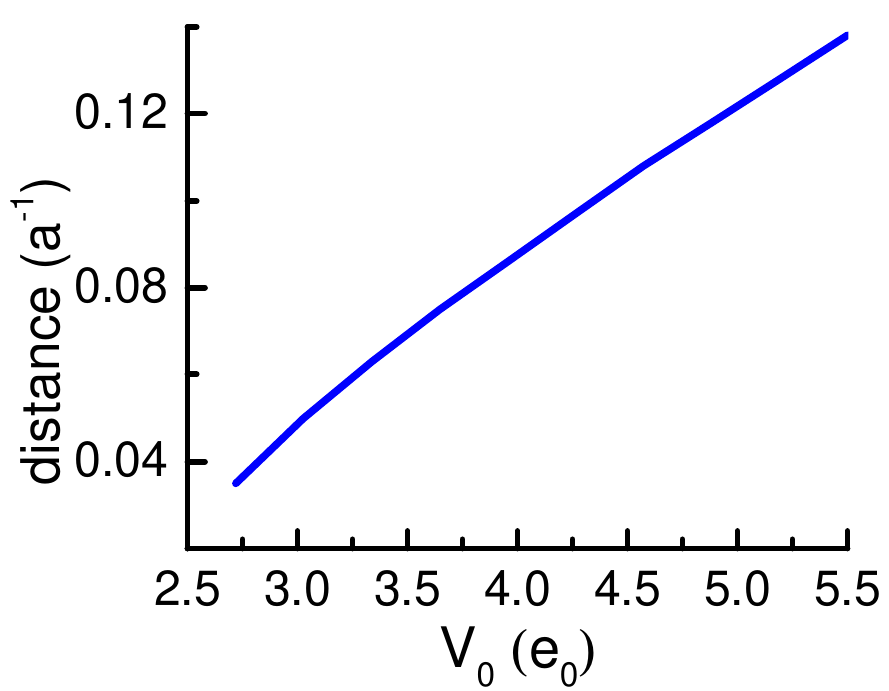}
  \caption{(Color online) The distance between original and new Dirac points as a function of $V_0$ in units of $e_0=\pi\frac{\hbar v_f}{L}$. $V_0$ are $2.405-5.52\pi\frac{\hbar v_f}{L}$, and in these $V_0$ there are three Dirac points in band structure. Other parameters are the same as Fig. \ref{fig1}.}\label{fig11}
\end{figure}

We also study the same properties when new Dirac points are far away from original one. We take $V_0=4\pi\frac{\hbar v_f}{L}$, $k_{0y}=0.03a^{-1}$, $0.117a^{-1}$, and $d=100a$. In this condition, Not all Dirac points are inside the wave packet. The Fig. \ref{fig10} shows ZB oscillations controlled by different Dirac points and they have clear distinctions. In Fig. \ref{fig10}(a), the center of wave packet is the closest to the original Dirac point, so the oscillation described by the red curve has the largest amplitude. In Fig. \ref{fig10}(b), the oscillation described by the blue curve has the largest amplitude, since the wave packet is the closest to 1st right new Dirac point so it is the main oscillation. In this condition, if we superpose these oscillation, it is equivalent to add a perturbation to the largest oscillation. So the oscillation is controlled by the nearest Dirac point mainly if not all Dirac points are inside wave packet. In Fig. \ref{fig11}, we research the distance between original and new Dirac points as a function of $V_0$. With increasing $V_0$, the distance increases monotonously, so we should choose a small $d$ to make all Dirac points inside wave packet if there is a large $V_0$. However, if $d$ is not large enough, the ZB will not have eligible amplitude and period to be observed, so we can not get an oscillation controlled by all Dirac points in large $V_0$. Moreover, if we take a small $V_0$ to make Dirac points are close to each other, we can get three similar oscillation controlled by three points respectively like Fig. {\ref{fig8}. Then we make all Dirac points at the same side of wave packet to make the oscillation are along the same direction, and choose an appropriate $k_{0y}$ to get period and amplitude which are observable. We can get an oscillation that are superposed by these three oscillations with larger amplitude, which is beneficial to observe ZB in experiments.}

Before the end, we consider the ZB when band structure has more Dirac points. In Fig. \ref{fig1}(d), there are five Dirac points in band structure, and we can find that band structure is linear near 2nd new Dirac points. Therefore, ZB oscillation near these points will have similar properties with ZB near original Dirac point. Furthermore,the distance between different Dirac points in momentum space is large that it is hard to make all of them inside a wave packet, so we can not research ZB controlled by all points and ZB will be influenced by the nearest Dirac point mainly.

Previous research only focuses on ZB near original Dirac point and fermions' initial momenta are along $k_y$-direction. We concentrate on the influences of decreasing group velocities in both $x$- and $y$-directions and the existence of new Dirac points. We also express ZB oscillations along the potential wells and indicate that group velocities could influence ZB differently when initial momenta are along different directions. These effects are quite different from ZB near original Dirac point and not discussed previously. Our results could be helpful for experiment observation on ZB in graphene superlattice.

\section{conclusions}
In summary, we have studied the ZB oscillation of fermions near new Dirac points in graphene superlattice. Compared with the ZB near original Dirac point, ZB near new Dirac points can have some unique properties because of the different group velocities. Additionally, we can choose appropriate potential and wave packet to obtain ZB controlled by all Dirac points. However, if Dirac points are so far away in momentum space that all points cannot be inside a wave packet, the ZB oscillation will controlled by the nearest Dirac points mainly, and we have to choose a small width of wave packet to make all Dirac points inside it then cause low amplitude. We should adjust potential to make band structure have three Dirac points which are not far away to obtain ZB influenced by all Dirac points.

\section*{acknowledgments}
J. L., S. L. and T. M. were supported in part by NSCFs (grant Nos. 11774033 and 11334012). L.-G. Wang was supported by Zhejiang Provincial Natural Science Foundation of China under Grant No. LD18A040001, and the grant by National Key Research and Development Program of China (No. 2017YFA0304202); it was also supported by the National Natural Science Foundation of China (grants No. 11674284 and U1330203), and the Fundamental Research Funds for the Center Universities (No. 2017FZA3005). We also acknowledge the support from by the HSCC of Beijing Normal University, and the Special Program for Applied Research on Super Computation of the NSFC-Guangdong Joint Fund (the second phase).

\bibliography{referenceZB}

\end{document}